\documentclass[aps,prl,twocolumn,superscriptaddress,showpacs]{revtex4}
\usepackage{graphicx}
\begin{document}
\def\214{Sr$_2$RuO$_4$}
\def\327{Sr$_3$Ru$_2$O$_7$}
\def\Ti327{Sr$_3$(Ru$_{1-x}$Ti$_x$)$_2$O$_7$}
\def\4310{Sr$_4$Ru$_3$O$_{10}$}
\def\113{SrRuO$_3$}
\def\RPSeries{Sr$_{n+1}$Ru$_n$O$_{3n+1}$}
\def\YCoAl{Y(Co$_{1-x}$Al$_x$)$_2$}

\title{Competing magnetic fluctuations in \327 probed by Ti doping}
\author{J. Hooper}
\affiliation{Department of Physics, Tulane University, New Orleans, Louisiana 70118, USA}
\author{M.~H. Fang}
\affiliation{Department of Physics, Zhejiang University, Hangzhou 310027, P.~R. China}
\author{M. Zhou}
\author{D. Fobes}
\author{N. Dang}
\author{Z.~Q. Mao}
\email{zmao@tulane.edu}
\affiliation{Department of Physics, Tulane University, New Orleans, Louisiana 70118, USA}
\author{C.~M. Feng}
\author{Z.~A. Xu}
\affiliation{Department of Physics, Zhejiang University, Hangzhou 310027, P.~R. China}
\author{M.~H. Yu}
\author{C.~J. O'Connor}
\affiliation{Advanced Materials Research Institute, University of New Orleans, New Orleans, Louisiana 70148, USA}
\author{G.~J. Xu}
\author{N. Andersen}
\affiliation{Materials Research Department, Ris\o\ National Laboratory, Frederiksborgvej 399, DK 4000 Roskilde, Denmark}
\author{M. Salamon}
\affiliation{Department of Physics, University of Illinois at Urbana-Champaign, Urbana, Illinois 61801, USA}

\date{\today}

\begin{abstract}
We report the effect of nonmagnetic Ti$^{4+}$ impurities on the
electronic and magnetic properties of \327. Small amounts of Ti
suppress the characteristic peak in magnetic susceptibility near 16
K and result in a sharp upturn in specific heat. The metamagnetic
quantum phase transition and related anomalous features are quickly
smeared out by small amounts of Ti. These results provide strong
evidence for the existence of competing magnetic fluctuations in the
ground state of \327. Ti doping suppresses the low temperature
antiferromagnetic interactions that arise from Fermi surface
nesting, leaving the system in a state dominated by ferromagnetic
fluctuations.
\end{abstract}

 \maketitle The
Ruddlesden-Popper-type perovskite strontium ruthenates \RPSeries
show a great diversity of electronic and magnetic ground states.
\214 ($n$=1), the most widely studied member of the series, is an
unconventional superconductor with a spin-triplet pairing
\cite{MaenoRMP,Maeno_Original214_Nature,Nelson_Science}. \113
($n$=$\infty$) is an itinerant ferromagnet with a Curie temperature
of 160 K \cite{Longo_113}. The $n$=2 bilayered member, \327, has an
intermediate dimensionality between these two; its magnetic ground
state was originally identified as an exchange-enhanced paramagnet
\cite{IkedaPRB}. Moderate applied fields induce a metamagnetic
transition in this material, with the transition field ranging from
4.9 T (for $H\parallel ab$) to 7.9 T (for $H\parallel c$)
\cite{Perry327,Grigera327}.

The metamagnetism in \327 has generally been interpreted as a
field-tuned Stoner transition into a highly polarized magnetic
state; since there is no broken symmetry, this transition is
expected to be first order \cite{WohlfarthRhodes,YamadaPRB}. In the
phase diagram one will thus have a first-order phase boundary line
which terminates in a critical end point \cite{Millis327}; there is
considerable experimental evidence that the characteristic
temperature of this end point is close to zero, indicating the
presence of quantum criticality \cite{Grigera327}. A variety of
unusual features, including non-Fermi liquid behavior and a possible
novel phase at the critical point, have been observed in this
material
\cite{Perry327,Grigera_PhaseForm_Science,Green_PhaseBifurcation_PRL}.
This finding has generated a great deal of interest and opened new
routes to exploring the novel physics of quantum criticality.

Based on the metamagnetic features, we might expect that the
relevant magnetic correlations in \327 are predominantly
ferromagnetic in nature \cite{Millis327}. However, inelastic neutron
scattering measurements have revealed that the system is dominated
by 2D ferromagnetic (FM) fluctuations only at high temperatures,
crossing over to 2D antiferromagnetic (AFM) fluctuations for
temperatures below 20 K \cite{CapognaPRB}. Furthermore, recent
$^{17}$O-NMR measurements have suggested that these low-temperature
AFM fluctuations continue to be dominant at high fields, including
directly at the quantum critical point \cite{Kitagawa_327NMR_PRL}.
This suggests a more complicated magnetic ground state for \327, in
which band FM correlations and AFM correlations due to nesting
effects might be in competition. Thus, further experiments to
elucidate the low-temperature magnetic ground state are highly
desirable.

Ti doping has been used as an effective probe of magnetic
correlations in the single-layered ruthenate \214. There, Ti was
found to enhance the anisotropic, incommensurate AFM fluctuations,
ultimately giving rise to a spin-density wave ordering with the same
nesting wavevector seen in the undoped material
\cite{Braden_Ti214Neutron_PRL,Ishida_Ti214NMR_PRB,Minakata_214,Kikugawa_Ti214_PRL,Pucher_214}.
Thus, to further probe the magnetic correlations in \327, we have
investigated the effects of doping nonmagnetic Ti in \327. In
contrast to \214, small levels of Ti suppress the AFM correlations,
leaving the system in a state with strong FM fluctuations. This
provides strong evidence for the existence of competing magnetic
fluctuations in the ground state of pure \327.

We have grown a series of Ti-doped crystals \Ti327 using a
floating-zone technique; growth conditions were similar to those
previously reported for pure \327 \cite{Perry_327Growth_JCG}.
Crystals selected for the measurements were characterized by x-ray
diffraction and did not include any impurity phase of \214 or \4310.
Magnetization measurements were taken in a SQUID magnetometer and
specific heat was measured by a standard thermal relaxation method
(Quantum Design, Model PPMS). Resistivity measurements were
performed with a standard four-probe technique.

\begin{figure}
\includegraphics{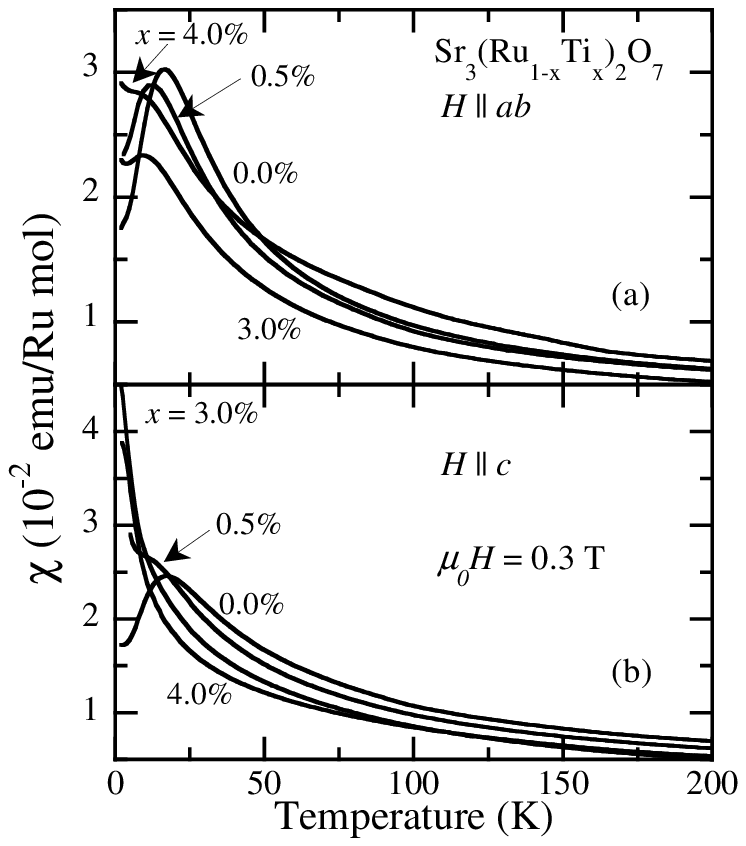}
    \caption{Temperature dependence of the uniform susceptibility $\chi$ = $M/H$ of Ti doped \327 for (a) in-plane and (b) out-of-plane fields, taken using an excitation field of $\mu_0H$ = 0.3 T.}
    \label{Ti327_Fig1}
\end{figure}

In Fig. \ref{Ti327_Fig1} we present the temperature dependence of
susceptibility $\chi$($T$) = $M/H$ for \Ti327 for (a) in-plane and
(b) out-of-plane field orientations. For the undoped sample, we
observe behavior similar to that previously reported by Ikeda
\textit{et al.} \cite{IkedaPRB}; susceptibility shows a peak around
16 K for both field orientations. The origin of this peak is likely
due to a crossover in the nature of magnetic correlations; inelastic
neutron scattering also shows a peak near 16 K for the dynamic
susceptibility $\chi^{\prime\prime}$ near a FM wavevector, while AFM
correlations increase rapidly below this temperature
\cite{CapognaPRB}. Recent NMR results further suggest that below
$\sim$16 K FM correlations are quenched and 2D incommensurate AFM
fluctuations become dominant \cite{Kitagawa_327NMR_PRL}.

We now consider the evolution of this feature in susceptibility
under Ti doping. For $H\parallel ab$ (Fig. \ref{Ti327_Fig1}a), Ti
impurities result in a supression of the peak near 16 K;
specifically, the low-temperature susceptibility increases until, by
4.0\% doping, only a small remnant of the peak remains. This trend
is even more remarkable for fields $H\parallel c$, shown in Fig.
\ref{Ti327_Fig1}b; by 4.0\% doping the peak is completely suppressed
and the susceptibility is reminiscent of a paramagnetic state. These
features suggest a change in the magnetic ground state such that 2D
AFM correlations are no longer dominant.

Above $\sim$180 K the susceptibility shows Curie-Weiss behavior.
Table \ref{Curie-Weiss fitting} shows the result of a fit to the
usual expression $\chi(T) = \chi_0 + C/(T-\Theta_W)$ for the samples
shown in Fig. \ref{Ti327_Fig1} for an in-plane field, where $\chi_0$
is the temperature independent term and $C/(T-\Theta_W)$ is the
Curie-Weiss term. The Weiss temperature $\Theta_W$ increases with Ti
doping, consistent with the idea that the system is moving towards a
FM ground state. However, the effective moment $p_{\text{eff}}$
derived from $C$ does not change remarkably with doping, which may
suggest that Ti substitution is acting primarily to suppress the low
temperature AFM correlations rather than directly enhancing the
ferromagnetism (see below for further discussion).

\begin{table}[b]
\caption{Parameters from Curie-Weiss fits, $H\parallel ab$
\label{Curie-Weiss fitting}}
\begin{ruledtabular}
\begin{tabular}{cccc}
$x$ & Curie constant \emph{C}& \emph{$\Theta_W$} & $p_{\text{eff}}$ \\
0.0 &   0.78622 &   -21.580 &2.5075\\
0.5 &   0.66488 & -12.287   &2.3059\\
3.0 &   0.58739 & 4.3060    &2.1673\\
4.0 &   0.66827 & 9.4349    &2.3117
\end{tabular}
\end{ruledtabular}
\end{table}

The enhanced Stoner model of itinerant metamagnets by Yamada gives
certain criterion for a first-order metamagnetic transition based on
a Landau expansion of the free energy \cite{YamadaPRB}.
Specifically, given the expansion $H = aM + bM^3 + cM^5$, where $H$
is the derivative of free energy with respect to the magnetization,
the ratio $ac/b^2$ should be less than 0.45 for a first-order
metamagnetic transition to occur. This ratio is related to the
susceptibility by the expression $ac/b^2 =
(5/28)[1-\chi(0)/\chi(T_{max})]^{-1}$ where $T_{max}$ is the
temperature where the peak in susceptibility occurs. For our
samples, this ratio is 0.423 for pure \327, 0.80 for 0.5\% doping,
and 6.13 for 3\% doping. While the magnetic properties are likely
more complex than a simple Stoner picture, this analysis suggests
that Ti doping quickly moves the system away from conditions
favorable for metamagnetism.

\begin{figure}
\includegraphics{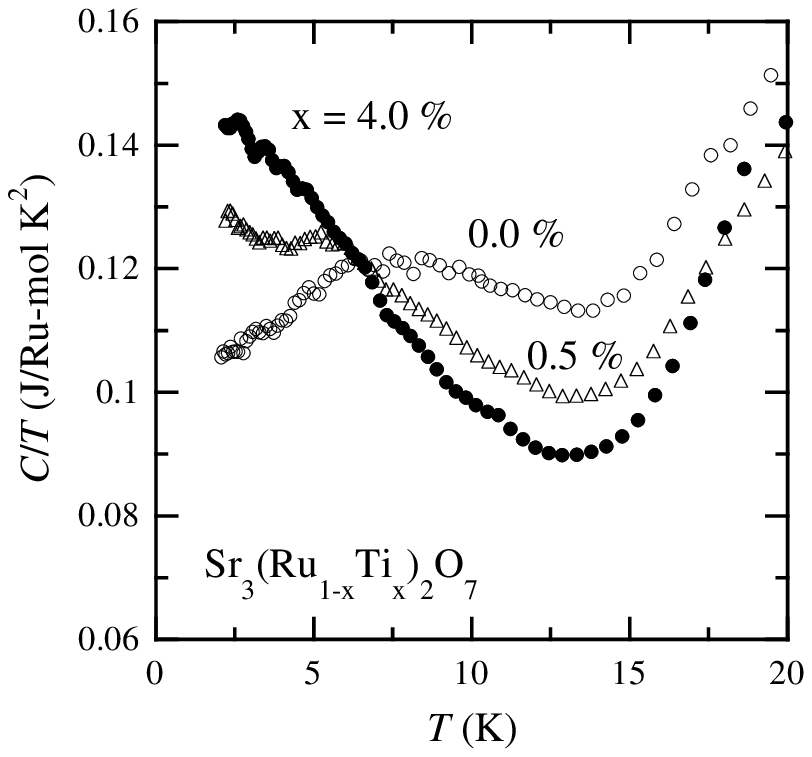}
    \caption{The specific heat divided by temperature for Ti doped \327 at low temperatures.}
    \label{Ti327_Fig2}
\end{figure}

Measurements of the specific heat give strong evidence that Ti
doping leaves the system in a ground state dominated by strong FM
fluctuations. In Fig. \ref{Ti327_Fig2} we present the specific heat
divided by temperature for the identical samples used in
magnetization measurements. All curves were taken at zero field. The
undoped sample is consistent with previously reported results by
Perry \textit{et al.} \cite{Perry327}; $C/T$ shows an upturn close
to 15 K, followed by a peak at lower temperatures. Ti doping
suppresses this low temperature peak, such that by 4\% doping we
observe a rapid increase in $C/T$ as temperature decreases.

While low temperature behavior consistent with a Schottky anomaly
has been previously reported in pure \327, it occurs only below 0.2
K and is likely not the source of the upturn observed here
\cite{Zhou_Trans327_PRB}. Under an applied magnetic field, $C/T$ in
in pure \327 also shows a sharp rise at low temperatures, consistent
with a log($T$) divergence due to proximity to a quantum critical
point (QCP) \cite{Perry327, Zhou_Trans327_PRB}. In our case we have
observed no other features which might suggest the presence of a
doping-induced QCP; rather, we believe the upturn in specific heat
arises from strong FM fluctuations. Further discussion of the origin
of this feature is given below.

\begin{figure}
\includegraphics{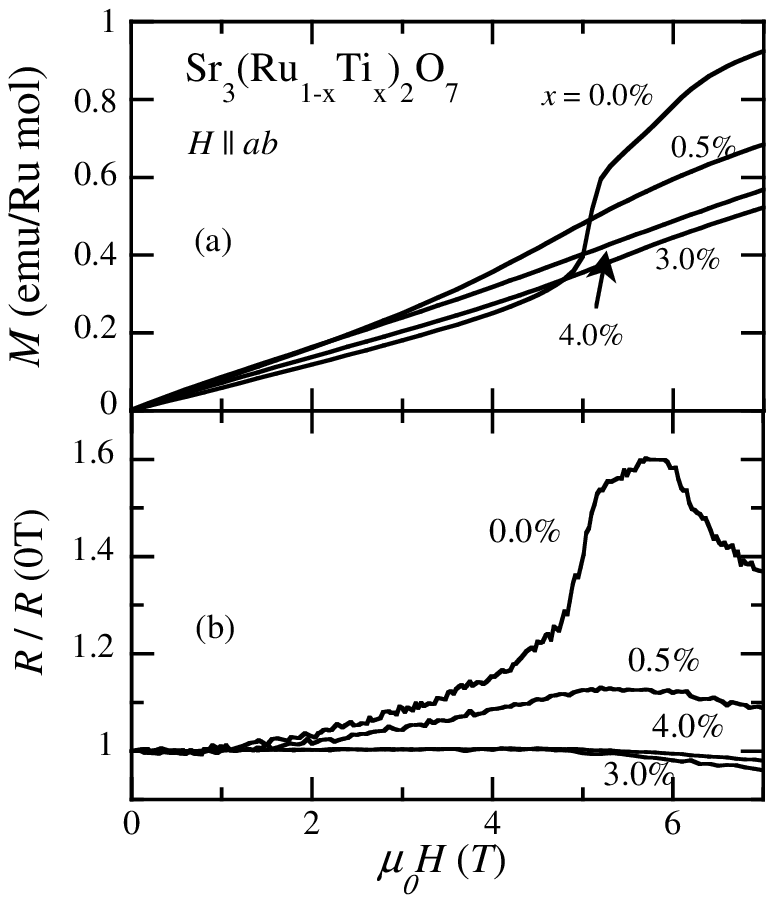}
    \caption{(a) Magnetization versus field and (b) resistance (normalized to its 0 T value) versus field for the \Ti327 system. Magnetization is taken at 2 K and resistance is measured at 0.3 K.}
    \label{Ti327_Fig3}
\end{figure}

We next consider how the features of the metamagnetic transition
change with doping. In Fig. \ref{Ti327_Fig3} we present the
magnetization and normalized resistance of Ti-doped \327 for
$H\parallel ab$. Consistent with previous reports, for the undoped
sample we observe a superlinear rise in magnetization and a broad
peak in resistance around the critical field. However, Ti impurities
act to rapidly suppress both of these features; even 0.5\% doping
has nearly smeared out the metamagnetic features in magnetization
and resistance. By doping levels of 3.0\% and 4.0\%, magnetization
is nearly linear in field and only a very small feature is seen in
resistance near 5 T. This observation is in agreement with Yamada's
criterion for metamagnetism to occur (see above).

\begin{figure}
\includegraphics{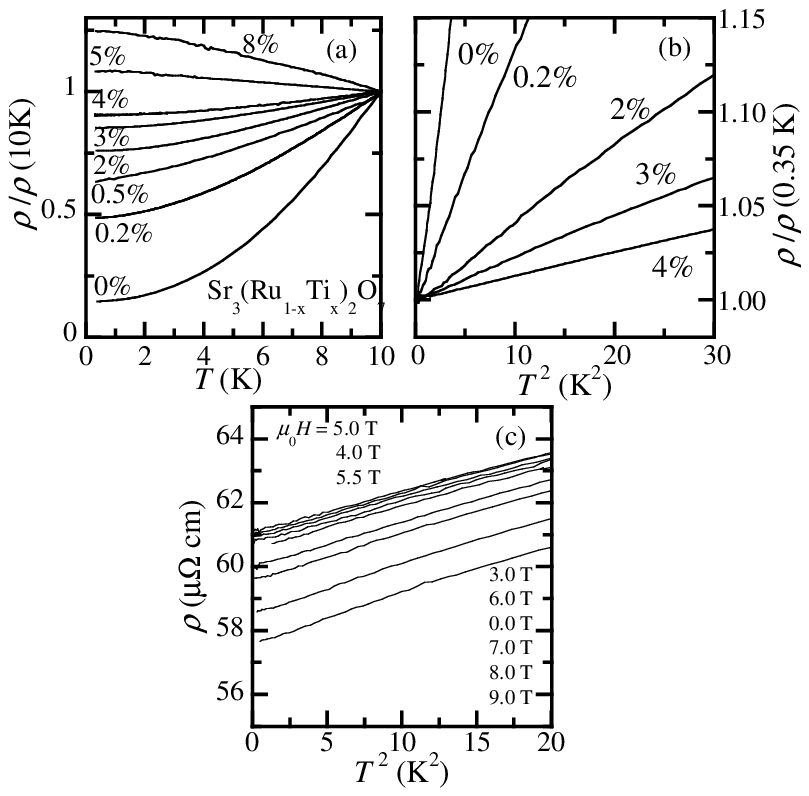}
    \caption{(a) The temperature dependence of resistivity for several Ti doping levels, normalized to the resistivity at 10 K. (b) Resistivity (normalized to its 0.3 K value) plotted versus $T^2$ for metallic doped samples. The data of the $x$= 0.5\% sample is not shown here, since it nearly overlaps with that of the pure sample. This is likely due to Ti inhomogeneity. (c) Resistivity of the 3.0\% doped sample versus $T^2$, under a range of applied fields. }
    \label{Ti327_Fig4}
\end{figure}

One of the most prominent features of the metamagnetic transition in
\327 is the non-Fermi liquid behavior close to the critical field,
manifested as a linear temperature dependence of resistivity
\cite{Perry327}. In Fig. \ref{Ti327_Fig4} we examine the electronic
ground state of Ti-doped \327 by means of bulk resistivity
measurements. Figure \ref{Ti327_Fig4}a shows the temperature
dependence of resistivity (normalized to its value at 10 K) for
several doping levels; the system remains metallic until
approximately 5\% doping, where it transitions into a localized
state. The metallic ground state remains a Fermi liquid, with
resistivity proportional to $T^2$ at low temperatures (see Fig.
\ref{Ti327_Fig4}b).

At a 3\% doping level there is still a small peak in the in-plane
susceptibility (see Fig. \ref{Ti327_Fig1}), but the metamagnetic
transition is almost completely suppressed and there is only a small
downturn in resistivity at 5 T. Figure \ref{Ti327_Fig4}c displays
the resistivity of this 3\% sample plotted against $T^2$ at various
applied fields for the configuration $H\parallel ab$. We can see
that the $T^2$ behavior of $\rho$ is unchanged by an applied field,
even close to where the metamagnetic transition occurs in the pure
sample; this indicates that quantum criticality is fully suppressed.

We now wish to discuss the origin of these features in \Ti327. We
begin by noting similarities between our results and previous
studies on other itinerant metamagnets; for example, the
intermetallic compound \YCoAl~ has a transition from an itinerant
metamagnetic state to a weakly FM one at a doping level of $x$ =
0.13 \cite{Yoshimura_OrigYCoAL_SSC}. For doping levels approaching
0.13, there is a suppression of the characteristic peak in
susceptibility \cite{Yoshimura_YCoAl_PRB}, a strong upturn in $C/T$
\cite{Wada_UpturnYCoAL_JPSJ}, and a significant broadening of the
metamagnetic transition \cite{Sakakibara_MMTYCoAl_PLA}. These
features have been successfully interpreted quantitatively within
the self consistent renormalized (SCR) theory of spin fluctuations
as evidence for proximity to a weakly FM state
\cite{Moriya_SpinFluc_JPSJ}. For \Ti327, as described above, Ti
doping results in similar changes as those seen in \YCoAl, clearly
suggesting that FM fluctuations become dominant.

In \YCoAl, the FM enhancement is believed to be associated with
doping-induced structural changes which increase the density of
states at the Fermi level \cite{Yoshimura_OrigYCoAL_SSC}. However,
in the case of \Ti327, the most likely mechanism is that both AFM
and FM fluctuations coexist in the ground state of the undoped
sample and that Ti doping primarily suppresses the AFM correlations.
We believe that the FM fluctuations are only slightly affected by
the presence of Ti, for the following reasons. First, at high
temperatures, where FM interactions are dominant, the susceptibility
and the effective moment do not change remarkably with increasing Ti
(see Fig. \ref{Ti327_Fig1} and Table \ref{Curie-Weiss fitting}).
Second, since Ti$^{4+}$ has a very similar ionic radius as
Ru$^{4+}$, substituting a small amount of Ti for Ru should have only
a very small effect on the structure; this is the case for Ti-doped
\214 \cite{Braden_Ti214Neutron_PRL}. All of these provide strong
support for the previous suggestion about the presence of competing
magnetic interactions in \327.

The existence of competing magnetic correlations in \327 is also
consistent with the results of band structure calculations
\cite{JapanBandStruct327,SinghMazin}. The bands derived from the Ru
$4d_{xy}$ orbitals are close to a van Hove singularity and thus
close to a FM Stoner instability, while bands derived from the
$4d_{xz}$ and $4d_{yz}$ orbitals have nesting features which would
favor AFM fluctuations \cite{CapognaPRB,SinghMazin}. The Ti doping
reported here appears to be be band-selective, primarily affecting
the character of the bands derived from $4d_{xz,yz}$ orbitals. This
is similar to Ti doping in the related compound \214, in which small
amounts of Ti have only minimal effects on the rotation of the
octahedra and the main $\gamma$ band, but do remarkably affect the
character of the 1D $\alpha$ and $\beta$ bands in such a way to
enhance the nesting features of the Fermi surface and thus enhance
the AFM correlations
\cite{Ishida_Ti214NMR_PRB,Braden_Ti214Neutron_PRL,Kikugawa_Ti214_PRL}.
In the case of \327 it appears that Ti impurities result in a
reduction of the nesting features, which is reasonable since the
characteristics of the bands involved in nesting are different
between the two materials \cite{SinghMazin214,SinghMazin}.

The disappearance of the metamagnetic features with Ti doping also
appears to suggest a possible relation between the short-range AFM
correlations and the metamagnetic transition. However, this
relationship may be complex; Ti doping levels of 5\% and above
result in a localized state as shown in Fig. \ref{Ti327_Fig4}a, and
this trend towards localization would drive the system away from the
Stoner instability, thereby reducing the metamagnetic features.
Thus, it is difficult from our data to draw direct conclusions about
the role of AFM correlations in the metamagnetic transition.

In conclusion, we have investigated the properties of Ti-doped \327.
We observe evidence that the short range AFM correlations are
quickly suppressed by small amounts of Ti, leaving the system in a
state dominated by 2D FM correlations. The most likely mechanism for
this is that Ti doping significantly alters the bands derived from
the Ru $4d_{xz,yz}$ orbitals, which results in a suppression of AFM
fluctuations due to nesting; however, it has less effect on the FM
correlations arising from the bands derived from the Ru $4d_{xy}$
orbitals. Ti doping thus confirms previous suggestions about the
presence of competing magnetic interactions in pure \327. This
result offers additional insight into the multiband nature of
magnetic correlations in this material.

We would like to thank A. Mackenzie, Y. Maeno, and Y. Liu for useful
discussions, and D. Niebieskikwiat for technical assistance. This
work was supported by the Louisiana Board of Regents support fund
LEQSF(2003-06)-RD-A-26 and pilot fund NSF/LEQSF(2005)-Pfund-23 at
Tulane, DARPA Grant No. MDA972-02-1-0012 at UNO, and the National
Science Foundation of China (No. 10225417) and the National Basic
Research Program of China (No. 2006CB601003) at Z.U. Z.Q.M. thanks
the Research Corporation for financial support.


\begin{thebibliography}{29}
\expandafter\ifx\csname
natexlab\endcsname\relax\def\natexlab#1{#1}\fi
\expandafter\ifx\csname bibnamefont\endcsname\relax
  \def\bibnamefont#1{#1}\fi
\expandafter\ifx\csname bibfnamefont\endcsname\relax
  \def\bibfnamefont#1{#1}\fi
\expandafter\ifx\csname citenamefont\endcsname\relax
  \def\citenamefont#1{#1}\fi
\expandafter\ifx\csname url\endcsname\relax
  \def\url#1{\texttt{#1}}\fi
\expandafter\ifx\csname urlprefix\endcsname\relax\def\urlprefix{URL
}\fi \providecommand{\bibinfo}[2]{#2}
\providecommand{\eprint}[2][]{\url{#2}}

\bibitem[{\citenamefont{Mackenzie and Maeno}(2003)}]{MaenoRMP}
\bibinfo{author}{\bibfnamefont{A.}~\bibnamefont{Mackenzie}} \bibnamefont{and}
  \bibinfo{author}{\bibfnamefont{Y.}~\bibnamefont{Maeno}},
  \bibinfo{journal}{Rev. Mod. Phys.} \textbf{\bibinfo{volume}{75}},
  \bibinfo{pages}{657} (\bibinfo{year}{2003}).

\bibitem[{\citenamefont{Maeno et~al.}(1994)}]{Maeno_Original214_Nature}
\bibinfo{author}{\bibfnamefont{Y.}~\bibnamefont{Maeno}} \bibnamefont{et~al.},
  \bibinfo{journal}{Nature} \textbf{\bibinfo{volume}{372}},
  \bibinfo{pages}{532} (\bibinfo{year}{1994}).

\bibitem[{\citenamefont{Nelson et~al.}(2004)}]{Nelson_Science}
\bibinfo{author}{\bibfnamefont{K.}~\bibnamefont{Nelson}} \bibnamefont{et~al.},
  \bibinfo{journal}{Science} \textbf{\bibinfo{volume}{306}},
  \bibinfo{pages}{1151} (\bibinfo{year}{2004}).

\bibitem[{\citenamefont{Longo et~al.}(1968)\citenamefont{Longo, Raccah, and
  Goodenough}}]{Longo_113}
\bibinfo{author}{\bibfnamefont{J.}~\bibnamefont{Longo}},
  \bibinfo{author}{\bibfnamefont{P.}~\bibnamefont{Raccah}}, \bibnamefont{and}
  \bibinfo{author}{\bibfnamefont{J.}~\bibnamefont{Goodenough}},
  \bibinfo{journal}{J. Appl. Phys.} \textbf{\bibinfo{volume}{39}},
  \bibinfo{pages}{1327} (\bibinfo{year}{1968}).

\bibitem[{\citenamefont{Ikeda et~al.}(2000)}]{IkedaPRB}
\bibinfo{author}{\bibfnamefont{S.}~\bibnamefont{Ikeda}} \bibnamefont{et~al.},
  \bibinfo{journal}{Phys. Rev. B} \textbf{\bibinfo{volume}{62}},
  \bibinfo{pages}{6089} (\bibinfo{year}{2000}).

\bibitem[{\citenamefont{Perry et~al.}(2001)}]{Perry327}
\bibinfo{author}{\bibfnamefont{R.~S.} \bibnamefont{Perry}}
  \bibnamefont{et~al.}, \bibinfo{journal}{Phys. Rev. Lett.}
  \textbf{\bibinfo{volume}{86}}, \bibinfo{pages}{2661} (\bibinfo{year}{2001}).

\bibitem[{\citenamefont{Grigera et~al.}(2001)}]{Grigera327}
\bibinfo{author}{\bibfnamefont{S.~A.} \bibnamefont{Grigera}}
  \bibnamefont{et~al.}, \bibinfo{journal}{Science}
  \textbf{\bibinfo{volume}{294}}, \bibinfo{pages}{329} (\bibinfo{year}{2001}).

\bibitem[{\citenamefont{Wohlfarth and Rhodes}(1962)}]{WohlfarthRhodes}
\bibinfo{author}{\bibfnamefont{E.}~\bibnamefont{Wohlfarth}} \bibnamefont{and}
  \bibinfo{author}{\bibfnamefont{P.}~\bibnamefont{Rhodes}},
  \bibinfo{journal}{Philos. Mag.} \textbf{\bibinfo{volume}{7}},
  \bibinfo{pages}{1817} (\bibinfo{year}{1962}).

\bibitem[{\citenamefont{Yamada}(1993)}]{YamadaPRB}
\bibinfo{author}{\bibfnamefont{H.}~\bibnamefont{Yamada}},
  \bibinfo{journal}{Phys. Rev. B} \textbf{\bibinfo{volume}{47}},
  \bibinfo{pages}{11211} (\bibinfo{year}{1993}).

\bibitem[{\citenamefont{Millis et~al.}(2002)}]{Millis327}
\bibinfo{author}{\bibfnamefont{A.~J.} \bibnamefont{Millis}}
  \bibnamefont{et~al.}, \bibinfo{journal}{Phys. Rev. Lett.}
  \textbf{\bibinfo{volume}{88}}, \bibinfo{pages}{217204}
  (\bibinfo{year}{2002}).

\bibitem[{\citenamefont{Grigera et~al.}(2004)}]{Grigera_PhaseForm_Science}
\bibinfo{author}{\bibfnamefont{S.~A.} \bibnamefont{Grigera}}
  \bibnamefont{et~al.}, \bibinfo{journal}{Science}
  \textbf{\bibinfo{volume}{306}}, \bibinfo{pages}{1154} (\bibinfo{year}{2004}).

\bibitem[{\citenamefont{Green et~al.}(2005)}]{Green_PhaseBifurcation_PRL}
\bibinfo{author}{\bibfnamefont{A.}~\bibnamefont{Green}} \bibnamefont{et~al.},
  \bibinfo{journal}{Phys. Rev. Lett.} \textbf{\bibinfo{volume}{95}},
  \bibinfo{pages}{086402} (\bibinfo{year}{2005}).

\bibitem[{\citenamefont{Capogna et~al.}(2003)}]{CapognaPRB}
\bibinfo{author}{\bibfnamefont{L.}~\bibnamefont{Capogna}} \bibnamefont{et~al.},
  \bibinfo{journal}{Phys. Rev. B} \textbf{\bibinfo{volume}{67}},
  \bibinfo{pages}{12504} (\bibinfo{year}{2003}).

\bibitem[{\citenamefont{Kitagawa et~al.}(2005)}]{Kitagawa_327NMR_PRL}
\bibinfo{author}{\bibfnamefont{K.}~\bibnamefont{Kitagawa}}
  \bibnamefont{et~al.}, \bibinfo{journal}{Phys. Rev. Lett.}
  \textbf{\bibinfo{volume}{95}}, \bibinfo{pages}{127001}
  (\bibinfo{year}{2005}).

\bibitem[{\citenamefont{Braden et~al.}(2002)}]{Braden_Ti214Neutron_PRL}
\bibinfo{author}{\bibfnamefont{M.}~\bibnamefont{Braden}} \bibnamefont{et~al.},
  \bibinfo{journal}{Phys. Rev. Lett.} \textbf{\bibinfo{volume}{88}},
  \bibinfo{pages}{197002} (\bibinfo{year}{2002}).

\bibitem[{\citenamefont{Ishida et~al.}(2003)}]{Ishida_Ti214NMR_PRB}
\bibinfo{author}{\bibfnamefont{K.}~\bibnamefont{Ishida}} \bibnamefont{et~al.},
  \bibinfo{journal}{Phys. Rev. B} \textbf{\bibinfo{volume}{67}},
  \bibinfo{pages}{214412} (\bibinfo{year}{2003}).

\bibitem[{\citenamefont{Kikugawa and Maeno}(2002)}]{Kikugawa_Ti214_PRL}
\bibinfo{author}{\bibfnamefont{N.}~\bibnamefont{Kikugawa}} \bibnamefont{and}
  \bibinfo{author}{\bibfnamefont{Y.}~\bibnamefont{Maeno}},
  \bibinfo{journal}{Phys. Rev. Lett.} \textbf{\bibinfo{volume}{89}},
  \bibinfo{pages}{117001} (\bibinfo{year}{2002}).

\bibitem[{\citenamefont{Minakata and Maeno}(2001)}]{Minakata_214}
\bibinfo{author}{\bibfnamefont{M.}~\bibnamefont{Minakata}} \bibnamefont{and}
  \bibinfo{author}{\bibfnamefont{Y.}~\bibnamefont{Maeno}},
  \bibinfo{journal}{Phys. Rev. B} \textbf{\bibinfo{volume}{63}},
  \bibinfo{pages}{R180504} (\bibinfo{year}{2001}).

\bibitem[{\citenamefont{Pucher et~al.}(2002)}]{Pucher_214}
\bibinfo{author}{\bibfnamefont{K.}~\bibnamefont{Pucher}} \bibnamefont{et~al.},
  \bibinfo{journal}{Phys. Rev. B} p. \bibinfo{pages}{104523}
  (\bibinfo{year}{2002}).

\bibitem[{\citenamefont{Perry and Maeno}(2004)}]{Perry_327Growth_JCG}
\bibinfo{author}{\bibfnamefont{R.}~\bibnamefont{Perry}} \bibnamefont{and}
  \bibinfo{author}{\bibfnamefont{Y.}~\bibnamefont{Maeno}}, \bibinfo{journal}{J.
  Crystal Growth} \textbf{\bibinfo{volume}{271}}, \bibinfo{pages}{134}
  (\bibinfo{year}{2004}).

\bibitem[{\citenamefont{Zhou et~al.}(2004)}]{Zhou_Trans327_PRB}
\bibinfo{author}{\bibfnamefont{Z.~X.} \bibnamefont{Zhou}} \bibnamefont{et~al.},
  \bibinfo{journal}{Phys. Rev. B} \textbf{\bibinfo{volume}{69}},
  \bibinfo{pages}{140409} (\bibinfo{year}{2004}).

\bibitem[{\citenamefont{Yoshimura and
  Nakamura}(1985)}]{Yoshimura_OrigYCoAL_SSC}
\bibinfo{author}{\bibfnamefont{K.}~\bibnamefont{Yoshimura}} \bibnamefont{and}
  \bibinfo{author}{\bibfnamefont{Y.}~\bibnamefont{Nakamura}},
  \bibinfo{journal}{Solid State Comm.} \textbf{\bibinfo{volume}{56}},
  \bibinfo{pages}{767} (\bibinfo{year}{1985}).

\bibitem[{\citenamefont{Yoshimura et~al.}(1988)}]{Yoshimura_YCoAl_PRB}
\bibinfo{author}{\bibfnamefont{K.}~\bibnamefont{Yoshimura}}
  \bibnamefont{et~al.}, \bibinfo{journal}{Phys. Rev. B}
  \textbf{\bibinfo{volume}{37}}, \bibinfo{pages}{3593} (\bibinfo{year}{1988}).

\bibitem[{\citenamefont{Wada et~al.}(1990)}]{Wada_UpturnYCoAL_JPSJ}
\bibinfo{author}{\bibfnamefont{H.}~\bibnamefont{Wada}} \bibnamefont{et~al.},
  \bibinfo{journal}{J. Phys. Soc. Jpn.} \textbf{\bibinfo{volume}{59}},
  \bibinfo{pages}{2956} (\bibinfo{year}{1990}).

\bibitem[{\citenamefont{Sakikibara et~al.}(1986)}]{Sakakibara_MMTYCoAl_PLA}
\bibinfo{author}{\bibfnamefont{T.}~\bibnamefont{Sakikibara}}
  \bibnamefont{et~al.}, \bibinfo{journal}{Phys. Lett. A}
  \textbf{\bibinfo{volume}{117}}, \bibinfo{pages}{243} (\bibinfo{year}{1986}).

\bibitem[{\citenamefont{Moriya and Kawabata}(1972)}]{Moriya_SpinFluc_JPSJ}
\bibinfo{author}{\bibfnamefont{T.}~\bibnamefont{Moriya}} \bibnamefont{and}
  \bibinfo{author}{\bibfnamefont{A.}~\bibnamefont{Kawabata}},
  \bibinfo{journal}{J. Phys. Soc. Jpn.} \textbf{\bibinfo{volume}{34}},
  \bibinfo{pages}{639} (\bibinfo{year}{1972}).

\bibitem[{\citenamefont{Hase and Nishihara}(1997)}]{JapanBandStruct327}
\bibinfo{author}{\bibfnamefont{I.}~\bibnamefont{Hase}} \bibnamefont{and}
  \bibinfo{author}{\bibfnamefont{Y.}~\bibnamefont{Nishihara}},
  \bibinfo{journal}{J. Phys. Soc. Jpn.} \textbf{\bibinfo{volume}{66}},
  \bibinfo{pages}{3517} (\bibinfo{year}{1997}).

\bibitem[{\citenamefont{Singh and Mazin}(2001)}]{SinghMazin}
\bibinfo{author}{\bibfnamefont{D.}~\bibnamefont{Singh}} \bibnamefont{and}
  \bibinfo{author}{\bibfnamefont{I.}~\bibnamefont{Mazin}},
  \bibinfo{journal}{Phys. Rev. B} \textbf{\bibinfo{volume}{63}},
  \bibinfo{pages}{165101} (\bibinfo{year}{2001}).

\bibitem[{\citenamefont{Mazin and Singh}(1999)}]{SinghMazin214}
\bibinfo{author}{\bibfnamefont{I.}~\bibnamefont{Mazin}} \bibnamefont{and}
  \bibinfo{author}{\bibfnamefont{D.}~\bibnamefont{Singh}},
  \bibinfo{journal}{Phys. Rev. Lett.} \textbf{\bibinfo{volume}{82}},
  \bibinfo{pages}{4324} (\bibinfo{year}{1999}).

\end{thebibliography}

\end{document}